\documentclass[11pt]{article}

\usepackage[final]{acl}

\usepackage{times}
\usepackage{latexsym}

\usepackage[T1]{fontenc}
\usepackage[utf8]{inputenc}

\usepackage{microtype}
\usepackage{inconsolata}
\usepackage{graphicx}
\usepackage{booktabs}
\usepackage{multirow}
\usepackage{amsmath,amssymb}
\usepackage{url}
\usepackage{todonotes}
\usepackage{tikz}
\usepackage{hyperref}
\usetikzlibrary{positioning, fit, backgrounds, calc}

\DeclareRobustCommand{\usericon}{\tikz[baseline=-0.6ex]{\fill[teal!90!black] (0,0.11) circle (2.2pt); \draw[line width=1.1pt, teal!90!black, cap=round] (-0.12,-0.11) .. controls (-0.09,-0.01) and (0.09,-0.01) .. (0.12,-0.11);}}

\title{Reproducing Transparent and Scrutable Recommendations: Exploring Open-Weight Models via Natural-Language User Profiles}

\author{
  Noah Mamie \\
  University of Zurich \\
  Department of Informatics \\
  \texttt{nmamie@ifi.uzh.ch}
  \And
  Laurin van den Bergh \\
  University of Zurich \\
  Department of Informatics \\
  \texttt{bergh@ifi.uzh.ch} \\
}

\begin{document}
\maketitle

\begin{abstract}
In this reproducibility study, we investigate the transparency and scrutability of recommender systems enhanced by incorporating generated natural-language user profiles that represent user preferences. The original paper explores the synthesis of user profiles from raw user-generated review text across domains such as movies and accommodations (Amazon Movies \& TV, TripAdvisor). Crucially, these natural-language user profiles enable direct user interaction and intervention, allowing users to customize recommendations by correcting misattributed preferences or addressing cold-start settings. We successfully reproduce the core findings of the original study. Additionally, we extend the evaluation by conducting systematic context ablation experiments, multi-seed stability across five distinct random seeds to establish statistical reliability, and a mechanistic interpretability analysis using the \texttt{nnsight} framework to probe internal model representations under counterfactual profile perturbations. Our findings verify the original paper's claim that User Profile Recommendation (UPR) achieves competitive performance under its test-set reranking protocol and makes recommendations more transparent. Perturbing the natural-language profiles does change predictions, but it shifts predicted ratings uniformly across genres with no detectable genre-selective effect, leaving rankings unchanged even under direct activation steering. We trace this back to the rating-regression objective rather than the profile interface, with ranking-objective models clearly exceeding in this task.

\end{abstract}

\section{Introduction}

Recommender systems have increasingly transitioned from the classical content-based and collaborative filtering paradigms to Large Language Model (LLM)-driven architectures \citep{deldjooReviewModernRecommender2024}. While classical approaches represent users as dense, uninterpretable latent vectors, recent work by \citet{ramos-etal-2024-transparent} introduces User Profile Recommendation (UPR). UPR synthesizes a user's past review history into a natural-language summary profile. This textual representation allows users to inspect, scrutinize, and directly modify their own profiles to customize future recommendations.

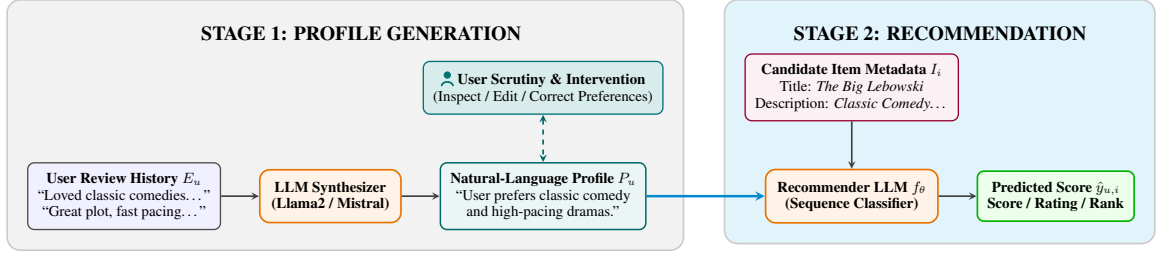
\begin{figure*}[t]
    \centering
    \resizebox{0.95\textwidth}{!}{%
    \begin{tikzpicture}[
        node distance=0.8cm and 1.1cm,
        box/.style={draw=black!70, fill=blue!6, rounded corners=4pt, align=center, font=\small, inner sep=6pt, line width=0.8pt},
        profilebox/.style={draw=teal!80!black, fill=teal!8, rounded corners=4pt, align=center, font=\small, inner sep=6pt, line width=0.9pt},
        itembox/.style={draw=purple!80!black, fill=purple!6, rounded corners=4pt, align=center, font=\small, inner sep=6pt, line width=0.8pt},
        llmbox/.style={draw=orange!90!black, fill=orange!10, rounded corners=6pt, align=center, font=\small\bfseries, inner sep=7pt, line width=1pt},
        outbox/.style={draw=green!70!black, fill=green!8, rounded corners=4pt, align=center, font=\small\bfseries, inner sep=6pt, line width=1pt},
        arrow/.style={->, >=stealth, line width=1pt, draw=black!75},
        dashedarrow/.style={<->, >=stealth, line width=1pt, dashed, draw=teal!80!black},
        stagebox/.style={rounded corners=8pt, inner sep=12pt, draw=black!20, line width=1pt}
    ]
        \node[box] (reviews) {\textbf{User Review History} $E_u$\\ \footnotesize ``Loved classic comedies\dots''\\ \footnotesize ``Great plot, fast pacing\dots''};
        
        \node[llmbox, right=0.8cm of reviews] (llm1) {\textbf{LLM Synthesizer}\\ \footnotesize (Llama2 / Mistral)};
        
        \node[profilebox, right=0.8cm of llm1] (profile) {\textbf{Natural-Language Profile} $P_u$\\ \footnotesize ``User prefers classic comedy\\ \footnotesize and high-pacing dramas.''};
        
        \node[box, above=1cm of profile, draw=teal!90!black, fill=teal!15] (user) {\usericon\ \textbf{User Scrutiny \& Intervention}\\ \footnotesize (Inspect / Edit / Correct Preferences)};
        
        \node[font=\large\bfseries, align=center, above=0.3cm of user, xshift=-10em] (stage1title) {STAGE 1: PROFILE GENERATION};
        
        \node[llmbox, right=2.5cm of profile] (recommender) {\textbf{Recommender LLM} $f_\theta$\\ \footnotesize (Sequence Classifier)};
        
        \node[itembox, above=1cm of recommender] (item) {\textbf{Candidate Item Metadata} $I_i$\\ \footnotesize Title: \textit{The Big Lebowski}\\ \footnotesize Description: \textit{Classic Comedy\dots}};
        
        \node[outbox, right=0.8cm of recommender] (output) {\textbf{Predicted Score} $\hat{y}_{u,i}$\\ \footnotesize Score / Rating / Rank};
        
        \node[font=\large\bfseries, align=center, xshift=5em] (stage2title) at (stage1title -| recommender) {STAGE 2: RECOMMENDATION };
        
        \begin{scope}[on background layer]
            \node[stagebox, fill=gray!10, fit=(reviews) (llm1) (profile) (user) (stage1title)] (stage1) {};
            \node[stagebox, fill=cyan!10, fit=(item) (recommender) (output) (stage2title)] (stage2) {};
        \end{scope}

        \draw[arrow] (reviews) -- (llm1);
        \draw[arrow] (llm1) -- (profile);
        \draw[dashedarrow] (user) -- (profile);
        
        \draw[arrow, line width=1.5pt, draw=cyan!70!blue] (profile) -- (recommender);
        
        \draw[arrow] (item) -- (recommender);
        \draw[arrow] (recommender) -- (output);

    \end{tikzpicture}%
    }
    \caption{The UPR architecture pipeline by \citet{ramos-etal-2024-transparent}, including our modifications to test different context ablations. User review histories $E_u$ are synthesized by an LLM into human-readable natural-language profiles $P_u$. Users can inspect and edit their profiles (scrutability/intervention). For user $u$ and item $i$, a fine-tuned sequence classification model $f_\theta$ predicts recommendation scores $\hat{y}_{u, i}$ given the profile $P_u$ and candidate item metadata $I_i$.}
    \label{fig:upr_architecture}
\end{figure*}

In this work, we present a comprehensive reproduction and extension of the UPR methodology. We evaluate the framework on the Amazon Movies \& TV \citep{he2016ups} and TripAdvisor\footnote{\href{https://www.tripadvisor.com}{https://www.tripadvisor.com}} \citep{li2021personalized} datasets, investigating both the recommendation accuracy and the semantic properties of the generated user profiles.\footnote{We release the \href{https://github.com/nmamie/transparent_user_profiles}{full code for optimal reproducibility}. We contacted the original authors to make them aware of this reproducibility study and to list points where the repository accompanying the original paper can be improved. They offered to consider our pull request to \href{https://github.com/jeromeramos70/user-profile-recommendation}{their repository.}}

Our primary contributions are as follows:
\begin{description}
    \item[Exact Reproduction \& Codebase Fixes:] We reproduce the original paper's primary experimental results, resolving several issues and inconsistencies in the original repository.
    \item[End-to-End Reproducible Pipeline:] We provide a fully reproducible codebase locked with \texttt{uv} dependency management, standardized CLI interfaces, and hardware-agnostic execution (\texttt{cuda}, \texttt{mps}, \texttt{cpu}).
    \item[Protocol Alignment \& Comparison:] We evaluate non-LLM baselines in Cornac \citep{salah2020cornac} under standardized evaluation protocols matching the LLM evaluation script, presenting side-by-side benchmark tables comparing original published numbers with our reproduction.
    \item[Context Ablations \& Multi-Seed Stability:] We conduct extensive context ablation experiments (comparing user profiles, raw review history, and item-review history) and evaluate model variance across five distinct random initialization seeds.
    \item[Mechanistic Interpretability \& Scrutability:] We leverage the \texttt{nnsight} interpretability library to probe internal model activations, map profile embeddings via UMAP, score cluster key terms via gradient attributions, and track latent representation trajectories under counterfactual user profile perturbations.
\end{description}

\section{Background \& Methodology}

\subsection{Problem Formulation}
Let $\mathcal{U}$ denote the set of users and $\mathcal{I}$ denote the set of items. Each user $u \in \mathcal{U}$ has an associated interaction history $E_u = \{e_{u,1}, e_{u,2}, \dots, e_{u,|E_u|}\}$, where $e_{u,i}$ represents a written review for an item $i$ previously consumed by $u$, and is often considered an explanation of a rating $y_{u,i}$. The UPR approach consists of two primary stages depicted in Figure~\ref{fig:upr_architecture}:

\begin{enumerate}
    \item \textbf{Profile Generation:} An LLM (e.g., Llama 2-7B or Mistral-7B) synthesizes the review history $E_u$ into a concise, natural-language user profile $P_u$:
    \begin{equation}
        P_u = \text{LLM\_Gen}(E_u)
    \end{equation}
    \item \textbf{Recommendation / Rating Prediction:} A sequence classification model $f_\theta$ is fine-tuned to predict item relevance $\hat{y}_{u,i}$ given the user profile $P_u$ and the item metadata $I_i$:
    \begin{equation}
        \hat{y}_{u,i} = f_\theta(\text{Prompt}(P_u, I_i))
    \end{equation}
\end{enumerate}

\subsection{User Scrutability \& Interventions}
Unlike black-box latent vector embeddings $\mathbf{h}_u \in \mathbb{R}^d$, the natural-language profile $P_u$ is human-readable. Users can edit $P_u$ directly by removing incorrectly inferred preferences (e.g., removing a dislike for a specific subgenre) or inserting new preference constraints (e.g., adding an interest in horror movies). 

\section{Scope of Reproduction \& Extensions}

\subsection{Scope of Reproduction}
Our reproduction focuses on validating the core claim of \citet{ramos-etal-2024-transparent}: that natural-language user profiles can match or exceed the predictive performance of raw review histories while maintaining human interpretability and scrutability. We reproduce fine-tuning of the recommender LLM $f_\theta$ and baseline evaluations on the Amazon Movies \& TV and TripAdvisor datasets using open-weight Llama2-7B \citep{touvron2023llama} and Mistral-7B \citep{jiang2023mistral} profile generations. We made sure to reuse the profiles generated by \citet{ramos-etal-2024-transparent}, to ensure an adequate and fair comparison.

\subsection{Extensions Beyond the Original Paper}
We extend the original study in three key directions:
\begin{enumerate}
    \item \textbf{Context Ablation Study:} We systematically evaluate model performance under three user representations (\texttt{user profile}, \texttt{review history}, and \texttt{item-review history}) paired with two item metadata representations (\texttt{item title} and \texttt{item title and description}).
    \item \textbf{Multi-Seed Robustness:} To ensure that reported gains are not artifacts of favorable random initializations, we conduct 5-seed stability trials on the Amazon dataset.
    \item \textbf{Mechanistic Interpretability \& Scrutability:} We leverage the \texttt{nnsight} interpretability library \citep{nnsight2024} to probe internal model activations, map profile embeddings via UMAP, score cluster key terms via gradient attributions, and track latent representation trajectories under counterfactual user profile perturbations.
\end{enumerate}

\section{Experimental Setup}

\begin{table*}[t]
    \centering
    \small
    \caption{Recommendation performance on Amazon Movies \& TV and TripAdvisor: Side-by-side comparison between original published numbers \citep{ramos-etal-2024-transparent} in Part (a) and our reproduced results in Part (b). Best value per section per dataset is in \textbf{bold}, and second-best is \underline{underlined}. Grayed-out entries ($^\dagger$) in Part (a) denote baselines whose source code was omitted from the official repository and could not be reproduced.}
    \label{tab:baseline_comparison}
    \begin{tabular}{lcccccccc}
    \toprule
    & \multicolumn{4}{c}{\textbf{Amazon Movies \& TV}} & \multicolumn{4}{c}{\textbf{TripAdvisor}} \\
    \cmidrule(lr){2-5} \cmidrule(lr){6-9}
    \textbf{Model} & \textbf{RMSE} $\downarrow$ & \textbf{MAE} $\downarrow$ & \textbf{nDCG@10} $\uparrow$ & \textbf{MAP} $\uparrow$ & \textbf{RMSE} $\downarrow$ & \textbf{MAE} $\downarrow$ & \textbf{nDCG@10} $\uparrow$ & \textbf{MAP} $\uparrow$ \\
    \midrule
    \multicolumn{9}{l}{\textit{\textbf{(a) Original Paper Reported Results (Ramos et al., 2024)}}} \\
    MostPop & 1.505 & 0.962 & 0.896 & 0.815 & 1.277 & 0.902 & 0.932 & 0.857 \\
    UserKNN & 0.960 & 0.707 & 0.935 & 0.864 & 0.836 & 0.634 & 0.952 & 0.883 \\
    Item-KNN & 1.045 & 0.790 & 0.897 & 0.823 & 0.890 & 0.683 & 0.953 & 0.852 \\
    MF & \textbf{0.925} & 0.686 & \textbf{0.941} & \textbf{0.870} & \textbf{0.786} & \textbf{0.599} & \textbf{0.960} & \textbf{0.891} \\
    NeuMF & 0.943 & 0.694 & 0.936 & 0.866 & 0.819 & 0.634 & 0.955 & 0.886 \\
    \color{gray}{PETER+$^\dagger$} & \color{gray}{\textbf{0.924}} & \color{gray}{0.685} & \color{gray}{\textbf{0.941}} & \color{gray}{\textbf{0.870}} & \color{gray}{0.803} & \color{gray}{0.621} & \color{gray}{0.958} & \color{gray}{\underline{0.889}} \\
    \color{gray}{PEPLER-MLP$^\dagger$} & \color{gray}{0.925} & \color{gray}{\textbf{0.672}} & \color{gray}{\textbf{0.941}} & \color{gray}{\underline{0.869}} & \color{gray}{\underline{0.793}} & \color{gray}{\underline{0.606}} & \color{gray}{\underline{0.959}} & \color{gray}{\underline{0.889}} \\
    \midrule
    UPR (Llama2) & 0.944 & \underline{0.678} & 0.938 & 0.866 & 0.804 & 0.616 & 0.955 & 0.885 \\
    UPR (Mistral) & 0.941 & 0.679 & \underline{0.940} & \textbf{0.870} & 0.804 & 0.610 & 0.952 & 0.887 \\
    \midrule
    \multicolumn{9}{l}{\textit{\textbf{(b) Our Reproduced Results}}} \\
    MostPop & 1.505 & 0.962 & 0.898 & 0.797 & 1.277 & 0.902 & 0.930 & 0.833 \\
    UserKNN & \underline{0.960} & \underline{0.707} & \underline{0.942} & \underline{0.918} & 0.836 & 0.634 & 0.958 & 0.934 \\
    ItemKNN & 1.045 & 0.790 & 0.906 & 0.874 & 0.890 & 0.683 & 0.934 & 0.901 \\
    MF & \textbf{0.925} & \textbf{0.686} & \textbf{0.948} & \textbf{0.924} & \textbf{0.786} & \textbf{0.599} & \textbf{0.964} & \textbf{0.942} \\
    NeuMF & n/a & n/a & 0.903 & 0.867 & n/a & n/a & 0.937 & 0.905 \\
    \midrule
    \textbf{UPR (Original)} & 0.952 & 0.687 & 0.945 & 0.920 & \underline{0.809} & \underline{0.618} & \underline{0.961} & \underline{0.939} \\
    \bottomrule
    \end{tabular}
\end{table*}

\subsection{Datasets \& Preprocessing}
We utilize the preprocessed Amazon Movies \& TV dataset and the TripAdvisor dataset, both formatted into the standard \texttt{train.jsonl}, \texttt{validation.jsonl}, and \texttt{test.jsonl} splits. Comprehensive details on data splits, prompt templates, fine-tuning configurations, hardware, and evaluation protocols are provided in Appendix~\ref{sec:appendix_setup}.

\subsection{Baseline Models \& Protocol Alignment}
To ensure fair comparison, we evaluate several standard recommendation baselines using Cornac \citep{salah2020cornac}: \texttt{MostPop}, \texttt{UserKNN}, \texttt{ItemKNN}, \texttt{MF}, and \texttt{NeuMF}. 

We aligned the baseline evaluation script (\texttt{rec\_baselines.py}) with the LLM evaluation protocol (\texttt{evaluate.py}). Exact metric formulas and candidate pool definitions are detailed in Appendix~\ref{sec:appendix_setup}.

\section{Results \& Analysis}
\label{sec:results}

\subsection{Baseline Comparison \& UPR Reproduction}

Table~\ref{tab:baseline_comparison} presents a direct side-by-side comparison between the original published numbers from \citet{ramos-etal-2024-transparent} in Part (a) and our reproduced results in Part (b).
Comparing Part (a) and Part (b) demonstrates strong qualitative alignment: classical matrix factorization (\texttt{MF}) remains the strongest rating accuracy baseline across both published and reproduced benchmarks. Nevertheless, UPR models achieve highly competitive ranking metrics on TripAdvisor (nDCG@10: 0.961, MAP: 0.939) while providing transparent natural-language profiles.
We highlight several key methodological nuances regarding baseline reproduction in Part (b):
\paragraph{Non-reproducible Baselines} As indicated in gray ($^\dagger$) in Table~\ref{tab:baseline_comparison}, \texttt{PETER+} and \texttt{PEPLER-MLP} could not be reproduced because the official open-source repository contains no implementation code or execution scripts for these architectures. We assume that the metrics reported in the original paper were imported directly from prior literature.
\paragraph{NeuMF Model Mismatch} Cornac's \texttt{NeuMF} is the implicit-feedback variant: rating
values never enter its objective, and its scores correlate with held-out ratings at $r{=}0.017$, so no training length or rescaling reaches $0.943$. The published figure reflects a rating-prediction NeuMF; we therefore report only ranking metrics numerically and the rest as \texttt{n/a}.
\paragraph{Single UPR Variant Execution} Our objective is to validate the reproducibility of the core UPR methodology for recommending items from \texttt{User Profile} $\rightarrow$ \texttt{Title}), rather than to assess profiles (\text the performance of various LLMs for user profile generation. Therefore, we focus Part (b) on reproducing the primary UPR configuration based on the original Llama2-generated user profiles.

As shown in our subsequent context ablation study (Table~\ref{tab:evaluation_results}), the UPR methodology is not only reproducible but also further improves when incorporating richer item metadata (\texttt{Title + Description}) or context history, achieving up to 0.895 MAP and 0.934 nDCG@10.

\subsection{Context Ablation}

\begin{table}[t]
    \centering
    \footnotesize
    \setlength{\tabcolsep}{2.5pt}
    \caption{Context Ablation Results on Amazon Movies \& TV. Original configuration in \textit{italics}, best in \textbf{bold}, second-best \underline{underlined}.}
    \label{tab:evaluation_results}
    \begin{tabular}{llcccc}
    \toprule
    \textbf{User ($P_u$)} & \textbf{Item ($I_i$)} & \textbf{RMSE} $\downarrow$ & \textbf{MAE} $\downarrow$ & \textbf{nDCG} $\uparrow$ & \textbf{MAP} $\uparrow$ \\
    \midrule
    \textit{Profile} & \textit{Title} & \textit{1.062} & \textit{0.788} & \textit{0.928} & \textit{0.894} \\
    Profile & Title+Desc & 1.063 & 0.784 & 0.929 & \underline{0.895} \\
    Reviews & Title & \textbf{1.043} & \underline{0.776} & 0.928 & 0.894 \\
    Reviews & Title+Desc & \underline{1.055} & \textbf{0.763} & 0.926 & 0.891 \\
    Item-Rev & Title & 1.105 & 0.864 & \underline{0.934} & \textbf{0.902} \\
    Item-Rev & Title+Desc & 1.108 & 0.879 & \textbf{0.934} & \textbf{0.902} \\
    \bottomrule
    \end{tabular}
\end{table}

Table~\ref{tab:evaluation_results} presents the ablation results across all input $I_i$ and output context configurations on the Amazon Movies \& TV dataset.
The ablation results confirm the core hypothesis of \citet{ramos-etal-2024-transparent}. Concise natural-language user profiles achieve rating prediction errors (MAE 0.784, RMSE 1.063) and ranking performance (MAP 0.895, nDCG@10 0.929) that are highly competitive with full raw review histories (MAE 0.763, nDCG@10 0.926), while using substantially shorter prompt contexts and offering full human readability.

\subsection{Multi-Seed Stability}

\begin{table}[t]
    \centering
    \small
    \setlength{\tabcolsep}{2.5pt}
    \caption{Multi-seed evaluation results on Amazon Movies \& TV across seeds 37--41 vs baseline (Seed 42, in \textit{italics}). Best in \textbf{bold}, second-best \underline{underlined}.}
    \label{tab:seed_stability}
    \begin{tabular}{lcccc}
    \toprule
    \textbf{Seed} & \textbf{RMSE} $\downarrow$ & \textbf{MAE} $\downarrow$ & \textbf{nDCG@10} $\uparrow$ & \textbf{MAP} $\uparrow$ \\
    \midrule
    37 & 0.948 & 0.690 & 0.945 & 0.921 \\
    38 & 0.949 & \textbf{0.686} & \textbf{0.946} & 0.921 \\
    39 & \textbf{0.947} & 0.689 & 0.945 & 0.920 \\
    40 & 0.960 & 0.705 & 0.944 & 0.919 \\
    41 & \underline{0.949} & 0.690 & \textbf{0.946} & \textbf{0.921} \\
    \textit{42} & \textit{0.952} & \textit{\underline{0.687}} & \textit{0.945} & \textit{\underline{0.920}} \\
    \midrule
    $\mu\pm\sigma$ & 0.951$\pm$.005 & 0.691$\pm$.007 & 0.945$\pm$.001 & 0.920$\pm$.001 \\
    \bottomrule
    \end{tabular}
\end{table}

To verify that evaluation significance is not impacted by random initialization or seed variance, we evaluated model performance across five distinct random seeds (seeds 37--41) alongside the reproduction baseline (Seed 42). Table~\ref{tab:seed_stability} summarizes the results on the Amazon Movies \& TV dataset.

The metrics display tight variance ($\sigma_{\text{MAP}} = 0.0009$, $\sigma_{\text{nDCG}} = 0.0007$), demonstrating that performance gains are statistically robust and independent of random seed choices (see Appendix~\ref{sec:appendix_setup} for detailed seed setup).

\subsection{Mechanistic Interpretability}
\label{sec:mech_interp}

To investigate how the fine-tuned recommender internalizes textual user profiles and responds to user interventions, we probe the final normalized layer of the transformer backbone using \texttt{nnsight}, extracting mean-pooled hidden activation vectors $\mathbf{h}_u \in \mathbb{R}^d$ for every user profile prompt (Appendix~\ref{sec:appendix_interp}). This captures the model's representation state immediately before classification.

\begin{figure*}[t]
    \centering
    \includegraphics[width=\textwidth]{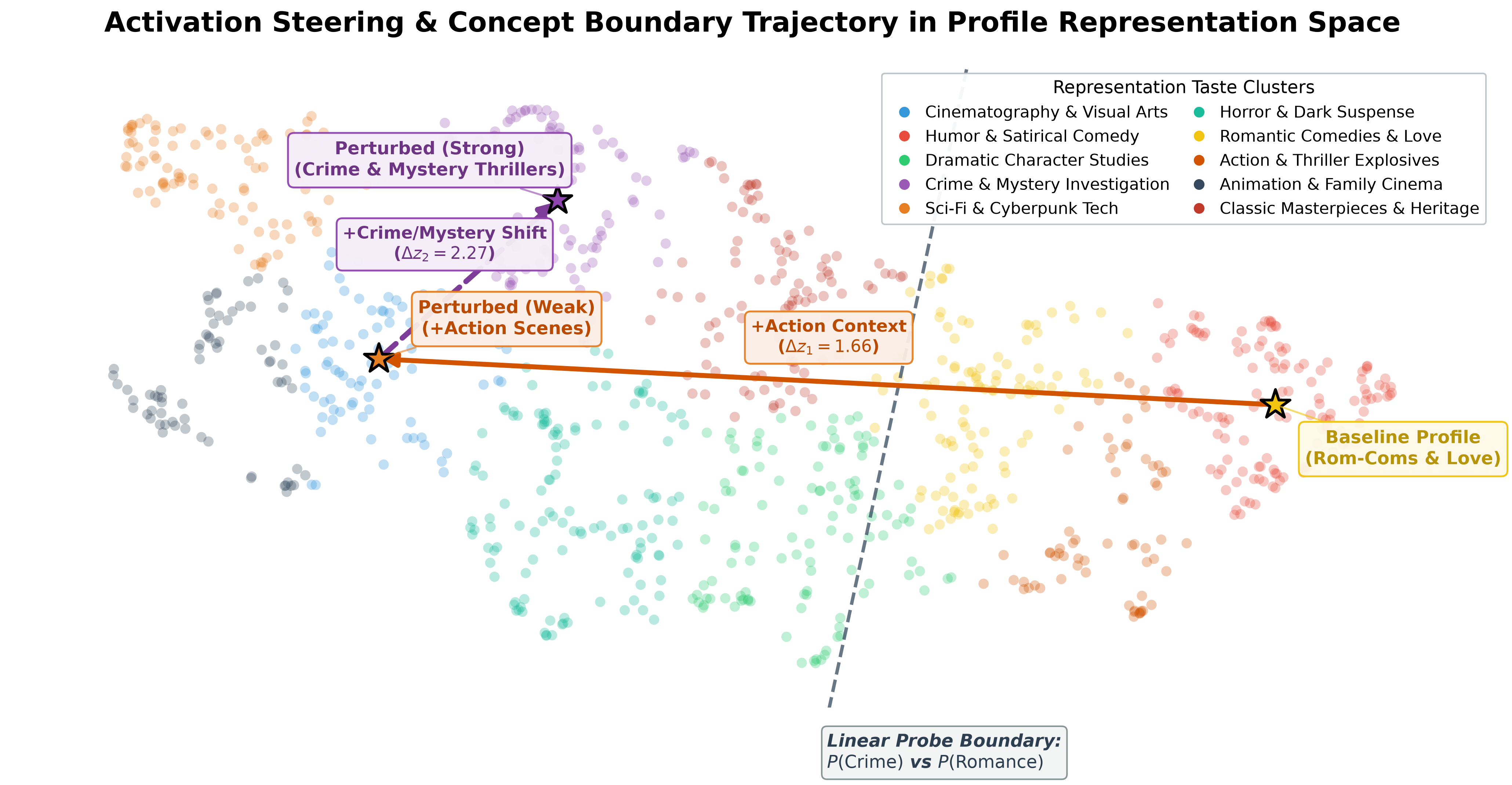}
    \caption{2D UMAP visualization of final layer normalization activations extracted via \texttt{nnsight}. Background point clouds are population-wide groupings of profile representations, color-coded and labeled by their most salient profile terms. Trajectory vectors show the three counterfactual edits of Table~\ref{tab:profile_perturbations}, with a fitted linear probe boundary.}
    \label{fig:perturbation}
\end{figure*}

\begin{table}[t]
    \centering
    \small
    \caption{Counterfactual profile perturbation states used in the trajectory analysis. Steering keywords added at each transition are in \textbf{bold}.}
    \label{tab:profile_perturbations}
    \begin{tabular}{@{}lp{3.3cm}l@{}}
    \toprule
    \textbf{State} & \textbf{Natural-Language Profile Text} & \textbf{Target} \\
    \midrule
    1 (baseline) & \textit{``I like romantic comedies with authentic love stories and lighthearted romance.''} & Romance \\
    2 (weak)     & \textit{``I like romantic comedies with \textbf{lots of action scenes and fast-paced thrillers}.''} & Action/Romance \\
    3 (strong)   & \textit{``My favorite genre is \textbf{crime and mystery thrillers with suspenseful plots, detective investigations, and dark puzzles}.''} & Crime \\
    \bottomrule
    \end{tabular}
\end{table}

\subsubsection{Representational structure}
\label{sec:repr_structure}

Figure~\ref{fig:perturbation} shows that natural-language profiles form well-separated groups in the model's 2D UMAP activation space, and tracks the representation under the three edits of Table~\ref{tab:profile_perturbations}. Re-embedding each edited profile yields smooth, continuous, and directional shifts: textual interventions do measurably change the model's internal user representation.

These groups do not align with user tastes. Labeling $1500$ users by the dominant genre of their \emph{actual} review history (romance $795$, crime $530$, action $175$; retained only when at least three reviewed items carry a genre and one genre holds the majority) and comparing against $K$-means over the same representations gives chance agreement: adjusted Rand index $0.0008$, normalized mutual information $0.0038$, with every cluster reproducing the population base rate ($46$--$59\%$ romance). A linear probe for the same label reaches $0.449$ accuracy against a $0.530$ majority-class baseline, above a shuffled-label control ($0.406$) but too weak to be useful. The clusters, therefore, organize profiles lexically, and the archetype labels describe profile language rather than validated taste segments.

\subsubsection{Do profile edits change predictions?}
\label{sec:repr_to_behavior}

Movement in representation space need not change recommendations. Table~\ref{tab:variance_decomposition} decomposes predicted-rating variance over a profile\,$\times$\,item grid. The profile is not inert, explaining $47.6$--$48.5\%$ of the variance, comparable to item identity, but it enters as a \emph{user-level rating offset} rather than a key matched against item content: the interaction term accounts for only $4.1\%$ with item titles alone and $16.1\%$ once descriptions are supplied. This bounds what any content-conditioned steering could explain, and all analyses below use the title-and-description configuration, since a title-only model never observes the genre evidence such an effect requires.

\begin{table}[t]
    \centering
    \small
    \caption{Variance decomposition of predicted ratings $\hat{y}$ over a grid of 25 real profiles $\times$ 24 catalog items. The profile explains roughly half the variance in both configurations, but as a user-level offset: the profile\,$\times$\, item interaction, where content-based personalization must live, accounts for $4.1\%$ and $16.1\%$ respectively.}
    \label{tab:variance_decomposition}
    \begin{tabular}{lccc}
    \toprule
    \textbf{Input Configuration} & \textbf{Item} & \textbf{Profile} & \textbf{Interaction} \\
    \midrule
    Profile + item title              & 48.3\% & 47.6\% & \phantom{0}4.1\% \\
    Profile + title \& descr.    & 35.5\% & 48.5\% & 16.1\% \\
    \bottomrule
    \end{tabular}
\end{table}

\begin{table*}[t]
    \centering
    \small
    \caption{Behavioral validation of the perturbation states in Table~\ref{tab:profile_perturbations}. For each state, we score $N{=}110$ catalog items per genre pool ($330$ candidates) with the profile\,+\, title\,\&\, description model, reporting mean predicted rating $\hat{y}$, mean rank in the pooled candidate set (lower is better), and the change relative to State~1.}
    \label{tab:behavioral_validation}
    \begin{tabular}{lcccccc}
    \toprule
     & \multicolumn{2}{c}{\textbf{State 1 (Baseline)}} & \multicolumn{2}{c}{\textbf{State 2 (Weak)}} & \multicolumn{2}{c}{\textbf{State 3 (Strong)}} \\
    \cmidrule(lr){2-3} \cmidrule(lr){4-5} \cmidrule(lr){6-7}
    \textbf{Item Genre Pool} & $\hat{y}$ & Rank & $\Delta\hat{y}$ & $\Delta$Rank & $\Delta\hat{y}$ & $\Delta$Rank \\
    \midrule
    Romance & 4.096 & 151.8 & $+0.186$ & $+1.9$ & $+0.059$ & $+0.0$ \\
    Action  & 4.008 & 167.8 & $+0.195$ & $-1.3$ & $+0.066$ & $+3.9$ \\
    Crime   & 3.999 & 176.9 & $+0.204$ & $-0.6$ & $\mathbf{+0.086}$ & $\mathbf{-3.9}$ \\
    \midrule
    \emph{Main effect (pooled)} & --- & --- & \multicolumn{2}{c}{$+0.195$, $p{<}10^{-50}$} & \multicolumn{2}{c}{$+0.070$, $p{<}10^{-9}$} \\
    \emph{Genre-selectivity of $\Delta\hat{y}$} & --- & --- & \multicolumn{2}{c}{$F{=}0.24$, $p{=}0.79$} & \multicolumn{2}{c}{$F{=}0.51$, $p{=}0.60$} \\
    \emph{Genre-selectivity of $\Delta$Rank}    & --- & --- & \multicolumn{2}{c}{$F{=}0.08$, $p{=}0.92$} & \multicolumn{2}{c}{$F{=}0.45$, $p{=}0.64$} \\
    \bottomrule
    \end{tabular}
\end{table*}

\paragraph{Textual edits.}
We build three genre pools (romance, action, crime) of $N{=}110$ catalog items each by weak supervision over item descriptions and score all $330$ candidates under each state. Editing the profile shifts predicted ratings substantially ($+0.195$ at State~2, $+0.070$ at State~3, both $p<10^{-9}$), but identically across the three pools (one-way ANOVA, $p{=}0.79$ and $p{=}0.60$; Table~\ref{tab:behavioral_validation}). The residual differences point the predicted way: under the crime-targeting edit, crime items gain most ($+0.086$ against $+0.059$ for romance) and are the only pool to rise in rank ($-3.9$ of $330$ positions, while action items fall by $3.9$). No contrast is significant (one-sided $p{=}0.15$ for $\Delta\hat{y}$, $p{=}0.31$ for $\Delta$Rank), the crime share of the top-$50$ moves only from $28\%$ to $30\%$, and a borderline result at $N{=}40$ did not survive tripling $N$.

\paragraph{Direct activation steering.}
A textual edit reaches behavior in two steps, and the above does not specify which step fails. We therefore intervene on the representation itself, adding $\alpha\mathbf{d}$ to the residual stream at the profile token positions, where $\mathbf{d}$ is the crime-minus-romance difference of means. Three checks precede any null: a zero vector and a \emph{constant} vector are both exact no-ops (layer normalization removes the latter, so steering directions must be non-uniform), while a random direction is not. Additive steering of the final representation cannot be selective at all, since the linear head shifts every item by the same $\alpha(\mathbf{w}\cdot\mathbf{d})$; we therefore steer at blocks 2, 6 and 9, early enough to interact with item tokens through attention.

The result dissociates representation from behavior (Table~\ref{tab:steering} in Appendix \ref{sec:appendix_interp}). The probe readout moves as intended, from $P(\text{crime}){=}0.064$ at $\alpha{=}0$ to $1.000$ at $\alpha{\geq}4$ -- a complete reversal of apparent genre preference -- while crime items' mean rating \emph{falls} slightly ($3.999 \rightarrow 3.984$) and their mean rank shifts by at most $0.2$ of $330$ positions at every layer. The probe reversal is partly guaranteed by construction, since the injected direction is the class-mean difference the probe reads; the behavioral null is the finding.

\subsubsection{Does the model personalize at all?}
\label{sec:personalization}

\begin{table}[t]
    \centering
    \small
    \caption{Percentile at which each ranker places a user's held-out positive item among $99$ items never interacted with ($200$ users, identical candidate sets; $0.5$ is chance).}
    \label{tab:personalization}
    \begin{tabular}{llcc}
    \toprule
    \textbf{Ranker} & \textbf{Objective} & \textbf{Percentile} & \textbf{Top-1} \\
    \midrule
    WMF                        & ranking    & $0.806$ & $13.0\%$ \\
    \texttt{MostPop}           & popularity & $0.697$ & \phantom{0}$6.0\%$ \\
    MF                         & rating     & $0.547$ & \phantom{0}$2.0\%$ \\
    UPR (title)      & rating     & $0.542$ & \phantom{0}$1.5\%$ \\
    UPR (title+desc) & rating     & $0.510^{\dagger}$ & \phantom{0}$1.5\%$ \\
    \bottomrule
    \multicolumn{4}{l}{\footnotesize $^{\dagger}$not different from chance ($p{=}0.62$); all others $p<0.05$.}
    \end{tabular}
\end{table}

These nulls admit a simpler explanation: there may be no personalization signal to steer. For each of the $200$ users, we rank one held-out positive item against $99$ items they have not interacted with, using identical never-interacted candidate sets for each ranker. The task is feasible on this data with a ranking-objective model, as Weighted Matrix Factorization (WMF) achieves $0.806$. The profile model reaches only $0.542$ with titles ($p{=}0.047$) and $0.510$ with descriptions ($p{=}0.62$), below \texttt{MostPop} ($0.697$), which personalizes not at all (Table~\ref{tab:personalization}). Its scores are uncorrelated with item popularity (Spearman's $r = -0.03$ to $-0.08$, $n.s.$), even though the positives are far more popular than the distractors ($81.6$ versus $41.9$ training interactions).

The ordering identifies the cause. While the ranking-objective model WMF reaches $0.806$, rating-regression objectives reach $0.51$--$0.55$. Every training pair is an item the user chose and reviewed, so the data holds no mismatched user--item pairs from which a content-matching rule could be learned, and the loss is minimized by a user offset plus an item offset -- the structure of Table~\ref{tab:variance_decomposition}. This is consistent with the metrics in \S\ref{sec:results}: test-set reranking over the Sakai condensed list orders only items the user already selected, for which item-side priors suffice. The limitation is thus a property of the rating-regression objective, inherited from the original formulation, rather than of natural-language profiles as an interface.

\subsubsection{Controls for the attribution analysis}
\label{sec:attribution_controls}

\begin{figure*}[t]
    \centering
    \begin{minipage}{0.48\textwidth}
        \centering
        \includegraphics[width=\linewidth]{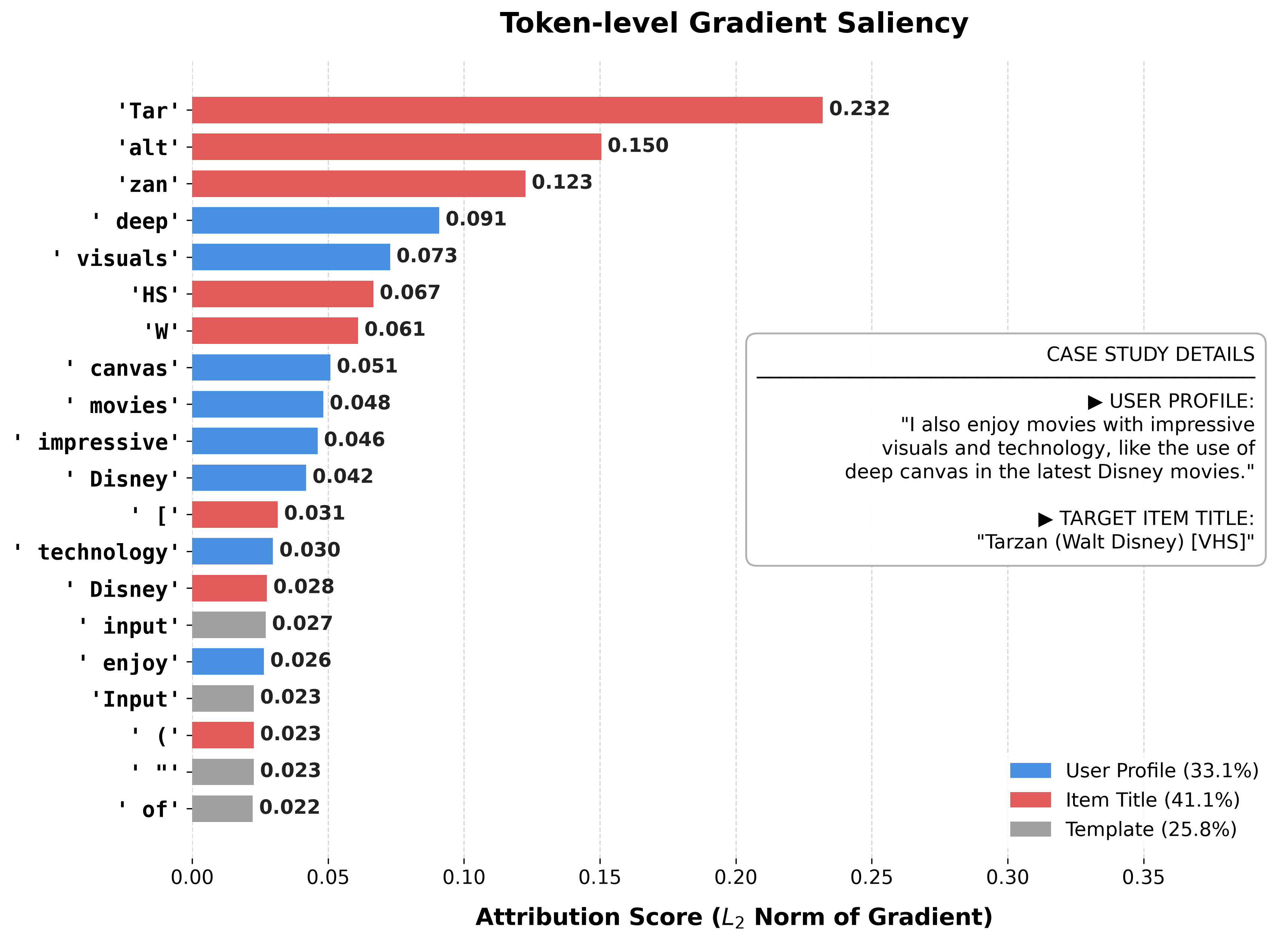}
        \caption{\label{fig:attribution}Local feature attribution for a single user profile and target item.}
    \end{minipage}\hfill
    \begin{minipage}{0.48\textwidth}
        \centering
        \includegraphics[width=\linewidth]{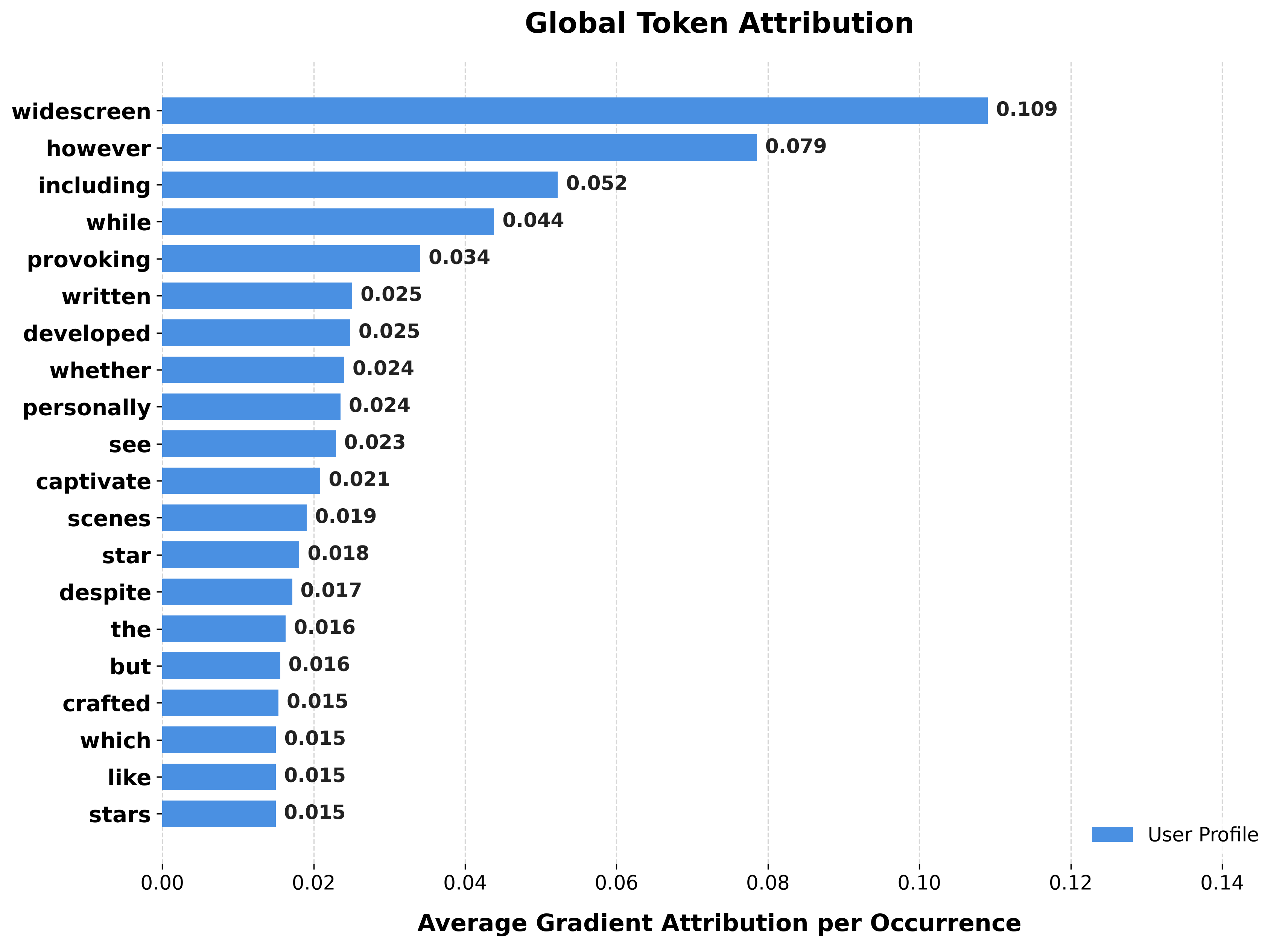}
        \caption{\label{fig:global_attribution}Global profile attributions over the Amazon user population.}
    \end{minipage}
\end{figure*}

Figures~\ref{fig:attribution} and~\ref{fig:global_attribution} report token-level gradient attribution over profile tokens. This could equally reflect the architecture, the pretrained backbone or the profile corpus, so we repeat the global analysis on the pretrained backbone with a randomly initialized head, on the same architecture without pretrained weights, and on word-shuffled profiles, comparing against corpus-frequency and subword-count baselines. Because gradient magnitudes are not comparable across models, we summarize each condition by a scale-invariant statistic: mean attribution on a fixed a priori genre lexicon divided by mean attribution on all other profile words.

The controls (Table~\ref{tab:attribution_controls} in Appendix \ref{sec:appendix_interp}) do not support reading these figures as evidence that fine-tuning concentrates attribution on preference features. Genre words receive roughly \emph{half} the attribution of other profile words (ratio $0.546$, $p{=}0.002$) and none appear in the top $20$, which is led by \texttt{widescreen} alongside discourse and style terms (\texttt{however}, \texttt{while}, \texttt{levity}, \texttt{cinematography}). Eleven of those $20$ recur under random initialization, and attribution correlates at $\rho{=}0.807$ with the fine-tuned model on word-shuffled profiles, so the pattern follows which words are present rather than their arrangement or meaning. Attribution is also negatively correlated with corpus frequency ($\rho{=}-0.320$, $p{=}0.027$), tracking word rarity, while subword count is not a significant driver ($\rho{=}0.212$, $n.s.$). We therefore treat these figures as descriptive of the saliency method and do not claim they isolate features acquired through fine-tuning.

Profile edits thus move the model's representation but not its recommendations, and scrutability in this formulation gives a user control over how generously they are modeled rather than what they are recommended. Training the same interface on a ranking objective with negative sampling is the change most likely to make the profile-edit behavior meaningful.

\section{Discussion on Reproducibility}

\paragraph{What was easy?}
The conceptual framework of UPR is intuitive, and the availability of pre-processed interaction datasets and initial source code provided a valuable starting point to reproduce the original paper's findings.

\paragraph{What was hard?}
Replicating the results required modifying the codebase to fix several issues:
\textbf{Syntax \& Variable Errors:} The original codebase contained a few broken variable references and undeclared identifiers.
\textbf{Missing Environment Definitions:} No \texttt{requirements.txt} or environment locks were supplied. We fixed this by introducing a clean \texttt{uv} environment specification.
\textbf{Evaluation Split Issue:} The original code validated the model using the test split instead of the validation split. Correcting this to ensure the test split is used only during testing yielded metric values closely aligned with the reported numbers.
\textbf{Baseline Protocol Realignment:} Original baseline scripts evaluated models under mismatched candidate protocols. Standardizing all baselines in \texttt{rec\_baselines.py} resolved this discrepancy.

\section{Conclusion}

This paper reproduces transparent and scorable recommendations using the UPR method developed by \citet{ramos-etal-2024-transparent} and extends it to new context configurations, random seeds for robustness checks, and mechanistic interpretability for probing the model activation space with counterfactuals. Our reproduction study confirms that the natural-language user profiles generated by UPR are both useful for the final recommendation task and increase the transparency and scrutability of the recommendation model. This enables users to better understand the recommendations provided to them and modify the suggested user profile, steering predictions towards a state that better reflects reality.

\section*{Limitations}

Our reproduction study focuses on the Amazon Movies \& TV and TripAdvisor datasets, and evaluates profiles synthesized via the Llama2-7B model. While our mechanistic interpretability analysis confirms structured feature representations in latent space, exploring profile stability under varied prompt templates and open LLM generation settings remains an important area for future research. Furthermore, our steering validation experiments reveal an important issue with the regression objective of the loss function, as it inherently diminishes the model's ability to rank items. A natural next step is to explore a ranking objective to assess whether the user-profile bottleneck allows meaningful steering of user predictions based on their interests.

\bibliography{custom}

\clearpage
\appendix
\raggedbottom

\section{Detailed Experimental Setup \& Hyperparameters}
\label{sec:appendix_setup}

To enable exact end-to-end reproduction, this section details the complete hardware, software, fine-tuning configuration, prompt formatting, model architectures, compute resources, and evaluation protocols used across all experiments.

\subsection{Software Environment \& Compute Resources}
All models were fine-tuned and evaluated using PyTorch 2.2+ and HuggingFace \texttt{transformers} 4.38+. Environment dependencies are strictly locked using \texttt{uv} via \texttt{pyproject.toml} and \texttt{uv.lock}.

\paragraph{Compute Resources \& Training Efficiency:}
Training a single UPR model configuration in its original baseline setup (\texttt{User Profile} $\rightarrow$ \texttt{Item Title}) on either the Amazon Movies \& TV or TripAdvisor dataset requires approximately \textbf{1 hour} of wall-clock training time when executed in parallel across \textbf{8 $\times$ NVIDIA RTX 4090 (24GB)} GPUs (using PyTorch DataParallel across device indices \texttt{CUDA\_VISIBLE\_DEVICES=0,1,2,3,4,5,6,7}). Computations were also validated for hardware-agnostic execution on \texttt{mps} (Apple Silicon) and \texttt{cpu}.

The full experiment scales proportionally based on the number of evaluated configurations:
\begin{itemize}
    \item \textbf{Context Ablation Study:} Evaluating all 6 combinations of input contexts (\texttt{user profile}, \texttt{review history}, \texttt{item-review history}) and output contexts (\texttt{item title}, \texttt{item title and description}) requires $6 \times$ single-run time ($\approx 6$ wall-clock hours on $8\times$ RTX 4090 GPUs).
    \item \textbf{Multi-Seed Stability Study:} Training across 5 distinct random initialization seeds (seeds 37--41) requires $5 \times$ single-run time ($\approx 5$ wall-clock hours on $8\times$ RTX 4090 GPUs).
\end{itemize}

\subsection{Fine-Tuning Configuration \& Hyperparameters}
In the codebase (\texttt{train.py}), recommendation scoring is framed as sequence classification fine-tuning. The model ($f_\theta$) is instantiated via HuggingFace's \texttt{AutoModelForSequenceClassification} initialized from pretrained causal LM backbones, defaulting to \texttt{gpt2} (124M parameters) or open-weight instruction models (\texttt{meta-llama/Llama-2-7b-chat-hf} and \texttt{mistralai/Mistral-7B-Instruct-v0.1}), with a single scalar output regression head (\texttt{num\_labels=1}).

\begin{itemize}
    \item \textbf{Model Initialization \& Tokenizer Setup:} 
    Loaded via HuggingFace's \texttt{AutoTokenizer}. Tokenizer padding is configured to right-side (\texttt{padding\_side="right"}) with sequence truncation enabled (\texttt{truncation=True}, \texttt{padding="max\_length"}), capped at a maximum sequence length of 300 tokens (\texttt{max\_length=300}). The padding token is explicitly mapped to the end-of-sequence token (\texttt{pad\_token = eos\_token}), and model config is aligned via \texttt{model.config.pad\_token\_id = model.config.eos\_token\_id}.
    \item \textbf{Target Normalization \& Inverse Transformation:} Ground-truth user rating scores $y \in [1, 5]$ are linearly normalized to $[0, 1]$ during dataset tokenization via $y_{\text{scaled}} = (y - 1)/4$. During evaluation, model scalar logit outputs $\hat{s} \in [0, 1]$ are transformed back to the original rating scale via $\hat{y} = \hat{s} \cdot 4 + 1$.
    \item \textbf{Optimizer, Learning Rate \& Precision:} Fine-tuning uses the AdamW optimizer with a linear learning rate decay schedule (\texttt{lr\_scheduler\_type="linear"}). The initial learning rate is set to $\eta = 3 \times 10^{-4}$ ($0.0003$). Computation runs in full precision (FP32 / FP16 depending on CUDA hardware capability).
    \item \textbf{Batching, Epochs \& Checkpoint Strategy:} Effective per-device batch size is set to 64 (\texttt{per\_device\_train\_batch\_size=64}, \texttt{per\_device\_eval\_batch\_size=64}). Fine-tuning runs for up to 5 epochs with epoch-wise evaluation and checkpointing (\texttt{save\_strategy="epoch"}, \texttt{eval\_strategy="epoch"}). Only the single best checkpoint is retained (\texttt{save\_total\_limit=1}, \texttt{load\_best\_model\_at\_end=True}). Early stopping is monitored via \texttt{EarlyStoppingCallback} with patience of 3 evaluation epochs (\texttt{patience=3}, \texttt{threshold=0.0}). Unused dataset columns are stripped during batch collation (\texttt{remove\_unused\_columns=True}).
    \item \textbf{Random Seeds:} Deterministic random initializations are enforced via \texttt{set\_random\_seeds(seed)} across PyTorch, NumPy, and Python \texttt{random}. Baseline reproduction uses seed 42, while multi-seed robustness tests evaluate seeds 37, 38, 39, 40, and 41.
\end{itemize}

\subsection{Evaluation Protocol \& Metric Definitions}
Evaluation is conducted via \texttt{evaluate.py} using ground-truth test splits for both Amazon and TripAdvisor Datasets. We used the original split sizes from \citet{ramos-etal-2024-transparent} to ensure a fair comparison:

\begin{itemize}
    \item \textbf{Test Candidate Pool:} For each user $u$, candidate items evaluated during ranking are restricted to the test interaction items in \texttt{test.jsonl}, aligning metric computation directly with standard test-set reranking protocols.
    \item \textbf{Calculated Metrics:}
    All metrics are implemented using the Cornac framework \citep{salah2020cornac}.
    \begin{enumerate}
        \item \textbf{RMSE (Root Mean Squared Error):}
        \begin{equation}
            \text{RMSE} = \sqrt{\frac{1}{N} \sum_{u,i} (\hat{y}_{u,i} - y_{u,i})^2}
        \end{equation}
        \item \textbf{MAE (Mean Absolute Error):}
        \begin{equation}
            \text{MAE} = \frac{1}{N} \sum_{u,i} |\hat{y}_{u,i} - y_{u,i}|
        \end{equation}
        \item \textbf{nDCG@10 (Normalized Discounted Cumulative Gain):} Measures ranking quality at cutoff $k=10$, normalized by the Ideal DCG (IDCG):
        \begin{equation}
            \text{DCG@k}(u) = \sum_{j=1}^{k} \frac{2^{y_{u,\pi(j)}} - 1}{\log_2(j + 1)},
        \end{equation}

        \begin{equation}
            \text{nDCG@k} = \frac{1}{|U|} \sum_{u \in U} \frac{\text{DCG@k}(u)}{\text{IDCG@k}(u)},
        \end{equation}
        
        where $\pi(j)$ represents the item ranked at position $j$ for user $u$.
        
        \item \textbf{MAP (Mean Average Precision):} Computes the mean of Average Precision (AP) across all users $U$:
        \begin{equation}
            \text{AP}(u) = \frac{1}{|R_u|} \sum_{j=1}^{N_u} P@j(u) \cdot \mathbb{I}(y_{u,\pi(j)} > 0),
        \end{equation}

        \begin{equation}
            \text{MAP} = \frac{1}{|U|} \sum_{u \in U} \text{AP}(u),
        \end{equation}
        
        where $R_u$ is the set of relevant items for user $u$, $P@j(u)$ is Precision at rank $j$, and $\mathbb{I}(\cdot)$ is an indicator function for relevance.
    \end{enumerate}
    \item \textbf{Output Aggregation:} Raw per-item prediction scores are saved, and aggregated dataset-wide metrics are written to disk.
\end{itemize}

\subsection{Prompt Construction \& Context Formats}
The model is fed structured natural-language prompts formatted as follows:

\paragraph{Prompt Template Format:}
Input context and candidate item metadata are formatted as:
\begin{verbatim}
Input Context: {input_context}
Based on the input context, from a 
scale of 1 to 5 (1 being lowest and 
5 being highest), I would give 
"{output_context}" a rating of
\end{verbatim}

Where the input and output contexts vary across our ablation configurations:
\begin{enumerate}
    \item \textbf{Input Context (\texttt{context\_in}):}
    \begin{itemize}
        \item \texttt{user profile}: The LLM-synthesized natural-language profile summary $P_u$.
        \item \texttt{review history}: Concatenation of up to 5 recent item reviews written by user $u$ excluding target item $i$.
        \item \texttt{item-review history}: Concatenation of up to 5 recent reviews written by other users for candidate item $i$.
    \end{itemize}
    \item \textbf{Output Context (\texttt{context\_out}):}
    \begin{itemize}
        \item \texttt{item title}: The candidate item title string $I_i$.
        \item \texttt{item title and description}: Candidate item title paired with text description ($I_i + D_i$).
    \end{itemize}
\end{enumerate}

\section{Mechanistic Interpretability Setup with \texttt{nnsight}}
\label{sec:appendix_interp}

To probe the internal hidden representations of fine-tuned recommendation LMs, we leverage the \texttt{nnsight} interpretability library:
\begin{itemize}
    \item \textbf{Probed Layer:} We trace hidden activations from the final layer normalization layer (\texttt{model.transformer.ln\_f}).
    \item \textbf{Pooling Method:} Mean-pooling across sequence tokens weighted by the input attention mask:
    \begin{equation}
        \mathbf{h}_u = \frac{\sum_{t=1}^T m_t \cdot \mathbf{h}_{u,t}}{\sum_{t=1}^T m_t}
    \end{equation}
    where $m_t \in \{0, 1\}$ is the attention mask at token position $t$.
    \item \textbf{UMAP \& Clustering:} High-dimensional representations $\mathbf{h}_u \in \mathbb{R}^d$ are projected into 2D UMAP space using cosine distance (\texttt{n\_neighbors=10}, \texttt{min\_dist=0.05}, \texttt{random\_state=42}). K-Means clustering ($K=10$) identifies preference archetype clusters.
    \item \textbf{Counterfactual Steering \& Profile Texts:} Three profile states (Baseline Romance, Weak Action, Strong Dark Crime; exact text in Table~\ref{tab:profile_perturbations}) are embedded and transformed through the fitted UMAP reducer to trace displacement vectors $\Delta d$ across a fitted linear probe decision boundary $P(\text{Romance})$ vs. $P(\text{Crime})$.
\end{itemize}

The following tables further support the analysis in \S\ref{sec:mech_interp}:

\begin{table}[t]
    \centering
    \small
    \caption{Activation steering at block 6. Adding $\alpha\mathbf{d}$ to the residual stream at the profile token positions, with $\mathbf{d}$ the crime-minus-romance difference of means, drives the probe readout across its full range while leaving ratings and rankings effectively unchanged. Blocks 2 and 9 behave identically. Rank is over the $330$-item set of Table~\ref{tab:behavioral_validation}.}
    \label{tab:steering}
    \begin{tabular}{lccccc}
    \toprule
    & \multicolumn{2}{c}{\textbf{Probe readout}} & \multicolumn{2}{c}{\textbf{Mean $\hat{y}$}} & \textbf{Crime} \\
    \cmidrule(lr){2-3} \cmidrule(lr){4-5}
    $\alpha$ & $P(\text{crime})$ & $P(\text{romance})$ & Crime & Romance & \textbf{rank} \\
    \midrule
    $0$ & 0.064 & 0.934 & 3.999 & 4.096 & 176.9 \\
    $1$ & 0.547 & 0.451 & 3.997 & 4.093 & 176.8 \\
    $2$ & 0.952 & 0.047 & 3.995 & 4.091 & 176.8 \\
    $4$ & 1.000 & 0.000 & 3.991 & 4.088 & 176.8 \\
    $8$ & 1.000 & 0.000 & 3.984 & 4.082 & 176.8 \\
    \bottomrule
    \end{tabular}
\end{table}

\begin{table*}[t]
    \centering
    \small
    \caption{Controls for the global attribution analysis. The genre ratio is mean attribution on a fixed genre lexicon over mean attribution on all other profile words ($1.0$ = no distinction). Agreement is the Spearman correlation of per-word attribution with the fine-tuned condition. Genre words are \emph{under}-attributed, and the pattern largely survives removing fine-tuning, removing pretraining, and shuffling word order.}
    \label{tab:attribution_controls}
    \begin{tabular}{lccc}
    \toprule
    \textbf{Condition} & \textbf{Genre ratio} & \textbf{Agreement} $\rho$ & \textbf{Top-20 overlap} \\
    \midrule
    Fine-tuned (reported)      & $0.546$ & ---      & --- \\
    Pretrained backbone only   & $0.381$ & $+0.452$ & $11/20$ \\
    Random initialization      & $1.169$ & $+0.229$ & $11/20$ \\
    Fine-tuned, shuffled words & $1.121$ & $+0.807$ & $10/20$ \\
    \bottomrule
    \multicolumn{4}{l}{\footnotesize vs corpus frequency: $\rho{=}-0.320$ ($p{=}0.027$); vs subword count: $\rho{=}0.212$ ($n.s.$).}
    \end{tabular}
\end{table*}

\end{document}